\documentstyle[aps,prl,twocolumn,epsf]{revtex}

\begin{document}

\twocolumn[\hsize\textwidth\columnwidth\hsize\csname @twocolumnfalse\endcsname

\draft

\title{Novel antiferromagnetic quantum phase transition in underdoped 
cuprates}

\author{Hyok-Jon Kwon\cite{emailjon}}
\address{Department of Physics, University of Florida, Gainesville,
FL 32611-8440}

\date{\today}

\maketitle

\begin{abstract} 
  We investigate a zero-temperature itinerant antiferromagnetic
transition where the fermions possess a $d$-wave gap. This problem
pertains to both the nodal liquid insulating phase and the $d$-wave
superconducting phase of the underdoped cuprates. We find that a
non-trivial quantum phase transition exists, and that the quantum
critical point is dominated by a long-ranged interaction ($|x-y|^{-2d}$)
of the N\'{e}el order parameter, which is induced by the Dirac-like
fermions near gap nodes. We formulate a Ginzburg-Landau functional and
estimate the critical exponents via the large-$n$ expansion method.

PACS numbers: 75.40.-s, 74.20.De, 75.50.Ee, 11.10.Hi
\end{abstract}

\vskip1pc 
]

Recent experiments  have revealed that underdoped cuprate materials exhibit
an exotic pseudogap phase under a characteristic temperature $T^*$.
There is ample evidence of the pseudogap phase above the superconducting
critical temperature\cite{pseudogap}; 
even a low-temperature pseudogap may exist in 
vortex cores in the mixed states\cite{sts}. 
Although the underlying microscopic mechanism of the pseudogap phenomena is
not understood, we may nevertheless study the
effective theory and its phenomenologies.
Based on the smallness of the  superfluid density in 
underdoped cuprates,
the pseudogap phase may be described as a superconductor whose
 phase coherence is destroyed by strong thermal phase fluctuations
but where the gap amplitude is robust\cite{Emery,qt}.  
As a possible zero-temperature analog of the pseudogap state, the idea of 
nodal liquid  has been 
proposed, as a superconductor which is quantum
disordered by  quantum phase fluctuations\cite{balents98}. The nodal
liquid is a non-magnetic insulating spin-liquid state,
which can be effectively described by charge-neutral fermionic 
quasiparticles with a $d$-wave gap, but with a non-zero insulating gap
in terms of the charge-ful electrons. 
This exotic state can be an intervening phase between the undoped 
antiferromagnetic (AF) insulating state and the the $d$-wave superconducting
state.
Further experimental and theoretical studies on the insulating 
and AF phases are 
needed to confirm this idea. 

In this report, a distinctive consequence of the nodal liquid state is 
presented:
We show that quasiparticles with
a $d$-wave gap cause a characteristic critical behavior of quantum phase 
transitions in the nodal liquid phase,
which is distinct from that of the conventional Fermi 
liquid\cite{hertz,millis93}. 
More specifically, we study an
itinerant AF transition at zero temperature. 
We expect, however, that without much modification
the same is true of certain classes of itinerant magnetic transitions
in a superconducting state of unconventional symmetry.
The origin of
the characteristic scaling is the long-ranged interaction between the
order parameter which is induced by massless Dirac-like fermions near  
gap nodes; after integrating out the fermionic degrees of freedom,
a term of the form ${\bf N}(x)\cdot{\bf N}(y)/|x-y|^{2d}$ is induced,
where ${\bf N}$ is the N\'{e}el order parameter, $d$ is the
spatial dimension, and $x,~y$ are $(d+1)$-dimensional space-time vectors. 
We show that a non-trivial quantum critical point exists at the transition
and that the abovementioned long-ranged interaction dominates the
critical behavior\cite{review1}.
The same argument holds without modification
for a zero-temperature AF transition
in  the $d$-wave superconductor although it is uncertain whether such a 
transition takes place in the {\it superconducting} state of cuprates.
This long-rangedness of the order parameter is similar to the case of 
itinerant ferromagnetic transition\cite{long-r} where non-critical soft modes
of the Fermi liquid induce such non-trivial interactions.
The difference is that in the case of Dirac fermions, additional soft modes 
are not needed to generate the
long-ranged order parameter interactions.


\begin{figure}[t]
\epsfxsize=4.5cm \epsfbox{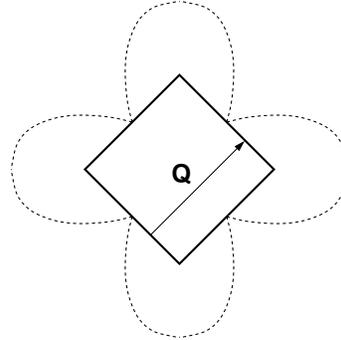}
\caption{
Schematic geometry of the Fermi surface and the $d$-wave gap.
The solid line is the nested Fermi surface, and the dashed line represents
the gap in the quasiparticle spectrum. $Q$ depicts an AF ordering wave-vector.
}
\label{FS}
\end{figure}

First, we construct an effective theory of nodal fermions with commensurate
AF exchange interaction. We will not discuss AF ordered phase in this paper.
Here and throughout we do not make a distinction 
between  quasiparticles in the $d$-wave superconducting state and  
charge-neutral nodal quasiparticles in the nodal liquid phase.
We consider a
system close to half-filling where
the gap nodes are located on the nested Fermi surface. (See Fig. \ref{FS}).
Since we are only interested in the low-energy properties,
we integrate out the fast momentum degrees of freedom and retain only 
the momenta near the Fermi surface.
Assuming that we can construct such a low-energy effective theory of the
Hubbard model near half-filling\cite{nest1}, 
we may express the effective action $S=S_0+S_{\rm AF}$
in the following form:
\begin{eqnarray}
S_0 &=&
T\sum_{\omega}\sum_{\bf k}
\left[
\psi^{\dag}_{\alpha}({\bf k};\omega)\left(i\omega-\xi_{\bf k}\right)
\psi^{\alpha}({\bf k};\omega) \right. 
\label{S0}
\\ 
&& \left. +\Delta^*_{\bf k}\psi^{\uparrow}({\bf k};\omega)~
\psi^{\downarrow}({\bf -k};-\omega)+ {\rm c.c.} \right]~, \nonumber \\
S_{\rm AF} &=& 
{1\over 2}J\int d\tau \sum_{\bf q}
{\bf n}_Q({\bf q},\tau )\cdot {\bf n}_Q({\bf -q},\tau )~,
\label{SAF}
\end{eqnarray}
where
${\bf n}_Q({\bf q},\tau )=\sum_{\bf k}{1\over 2}~\psi^{\dag}_{\alpha}
({\bf k},\tau)\vec{\sigma}^{\alpha}_{\beta}\psi^{\beta}({\bf k+Q_k+q},\tau)
$, and $\Delta_{\bf k}$ is a $d$-wave gap.
Here $|{\bf q}| \ll k_F$ and ${\bf Q_k}$ is a commensurate nesting
vector antiparallel to the Fermi velocity at ${\bf k}$. Also we take
$J <0$ so that an AF transition takes place. 

Before we discuss the AF critical point, we revisit results of the
mean-field theory.
To obtain a mean-field theory, it is sufficient to
employ a mixed representation of the momentum to separate the
Fermi surface label and the small relative momentum near the Fermi surface
as follows:
\begin{eqnarray}
S &=& S_0 +\int d^2 x~d\tau \bigg{[}
-{1\over 2J}~{\bf N}^2( x) \label{oAct}
 \\ 
&& -\sum_{\bf p}e^{-i{\bf Q_p\cdot x}}{\bf N}( x)\cdot
{1\over 2} \psi^{\dag}_{\alpha}({\bf p};x)~\vec{\sigma} ^{\alpha}_{\beta}
\psi^{\beta}({\bf p+Q_p};x)  \bigg{]} \nonumber
\end{eqnarray}
where 
\begin{eqnarray}
S_0 &=&
T\sum_{\omega}\sum_{\bf p,q} \left\{
\psi^{\dag}_{\alpha}({\bf p};q)\left[i\omega-\xi_{\bf q}({\bf p})\right]
\psi^{\alpha}({\bf p};q) \right. \label{fAct}
 \\ 
&& \left. +\Delta^*_{\bf p}\psi^{\uparrow}({\bf p};q)~
\psi^{\downarrow}({\bf -p};-q)+ {\rm c.c.} \right\}~. \nonumber
\end{eqnarray}
Here we have separated the AF exchange interaction by
introducing a staggered magnetization ${\bf N}$ via Hubbard-Stratonovich
transformation. The momentum ${\bf p}$ labels points on the Fermi surface,
and ${\bf q}$ is a small momentum deviation from the Fermi surface.
$q =({\bf q}, \omega)$ is a $(d+1)$-dimensional wave-vector and similarly,
 $x=({\bf x}, \tau)$.
In this representation, the $d$-wave symmetry of the gap dictates that
$\Delta_{\bf p}=\Delta_{\bf -p}$ and $\Delta_{\bf p}=-\Delta_{\bf p+Q_p}$.
Taking $|{\bf N}|=N_z$=constant, we obtain, for instance, the quasiparticle
Greens function
${\cal{G}}({\bf p};q)=-(i\omega+\xi_{\bf q})/( \omega ^2 +\xi^2_{\bf q}
+|\Delta_{\bf p}|^2+N^2/4 )$. As expected, the quasiparticle gains
a gap of the form $\sqrt{|\Delta_{\bf p}|^2+N^2/4 }$.

From the following mean-field self-consistency condition,
\begin{eqnarray}
{N\over |J|} &=&\sum_{\bf p}
{1\over 2}~(\sigma^z)^{\alpha}_{\beta}\langle
\psi^{\dagger}_{\alpha}({\bf p};x)~\psi^{\beta}({\bf p+Q_p};x)\rangle ~,
\end{eqnarray}
we obtain the relation
\begin{equation}
|N|/2+\sqrt{\Delta^2_0+N^2/4} = E_{\Lambda}e^{-1/|J|N_0}~,
\label{selfc}
\end{equation}
where $E_{\Lambda}$ is an  upper energy cutoff which is of order $E_F$,
 $N_0$ is the density of states at the Fermi surface, and $\Delta_0$
is the maximum gap.
This suggests that due to the gap in the quasiparticle spectrum,
the AF transition is inhibited to such a degree that
a sufficiently strong exchange interaction is needed for a transition.
In fact, we infer from Eq. (\ref{selfc}) that the AF coupling $J$ needs
to be at least of comparable strength to the superconducting pairing strength;
on a circular Fermi surface, even stronger exchange interactions would be
required\cite{Kee}.
This is in contrast to the case of Fermi liquid on a nested Fermi surface
where arbitrary strength of exchange interaction leads to an AF
instability\cite{nest1}. 
Therefore, a weak-coupling fermionic renormalization group (RG) analysis is 
inadequate for this problem. Instead, we turn to the 
Ginzburg-Landau (GL) functional of  the N\'{e}el order parameter below.

Now we come back to Eq. (\ref{S0}) which reduces to a spatially anisotropic
massless Dirac fermionic action near
the gap nodes. In terms of the small momentum deviation $\bf q$
from the Fermi wave-vector, the fermion  Greens functions have 
poles at $\omega \approx \pm i\sqrt{v_F^2q_{\parallel}x^2+v_{\Delta}^2
q_{\perp}^2} $,
where $v_F$ is the Fermi velocity, $v_{\Delta}$ is the slope of the
gap at the node in the momentum space, and $q_{\parallel}$ ($q_{\perp}$)
is the component of momentum parallel (perpendicular) to the Fermi
velocity at the node.
Although the bare theory does not have an exact Lorentz symmetry
due to this anisotropy, we assume that  even if we enforce Lorentz symmetry
by hand, the quality of the quantum critical point is not crucially 
changed\cite{balents98} 
and take $v_F=v_{\Delta}=1$ for simplicity.

To arrive at the GL functional of the order parameter, we
integrate out the fermionic degrees of freedom. For now we assume that this
procedure is under control, although we will show below that at $T=0$ it is
beset with singularities in GL coefficients.
We confine our discussion to an AF transition sufficiently far away from 
a superconductor-insulator transition, and retain only the N\'{e}el
order parameter.
After formally integrating out the fermionic fields,
\begin{eqnarray}
S[{\bf N}] &=&
\sum_n\int d^dx_1~d\tau_1...d^dx_n~d\tau_n~
\Gamma_n(x_1,...,x_n)\nonumber \\
&&\times {\bf N}(x_1)...{\bf N}(x_n)~,
\label{gl}
\end{eqnarray}
where $n$ is an even integer.
The Fourier transform of $\Gamma_2$ \cite{Kee} is obtained from
\begin{eqnarray}
\Gamma_2({\bf q},\nu) &=& -{1\over 2J} +{1\over 2}\int {d\omega \over 2\pi}
\sum_{\bf k}[{\cal G}({\bf k},\omega)~{\cal G}({\bf k+Q_k+q},\omega+\nu)
]~, \nonumber \\
&&~
\end{eqnarray}
where ${\cal G}$  is the diagonal  component of
the quasiparticle Greens function.
Collecting the lowest powers of the momentum and frequency, we find that 
for $2\le d < 3$, which is the range of dimensionality of our interest, the
leading term is non-analytic in $|q|$:
\begin{equation}
\Gamma_2=t+c|{ q}|^{d-1}~,
\label{key}
\end{equation}
where $|q|=\sqrt{{\bf q}^2 +\nu ^2}$ and $c>0$.
 Similarly, one can show that  $\Gamma_n \sim | q|^{d-n+1}$.
Using the Gaussian approximation, $\langle {\bf N}\cdot {\bf N}\rangle 
=1/(t+c_2| q|^{d-1}) $. Taking $t$ as the 
distance from the critical point, we obtain various critical exponents:
$\gamma = 1~,~ z=1~,~\nu=1/(d-1)$, and $\eta = 3-d$.
By including higher order GL expansions and introducing an
$O(3)$ symmetry-breaking
term ${\bf h\cdot N}$, we can obtain the equation of state.
In the presence of the staggered magnetic field, a non-zero $N$ value will
cut off the infrared singularities in $\Gamma_n $ by providing a low-momentum 
cutoff\cite{long-r}. Therefore, $\Gamma_n \sim N^{d-n+1}$ obtains, and the
equation of state ($\delta S/\delta N =0$) is $tN+N^d\approx h$. 
From this, we obtain two more exponents:
$\beta = 1/(d-1)$ and $\delta = d$.

Fixing the Gaussian term as marginal, we may estimate the scaling dimensions
of higher GL expansions. By transforming $q \rightarrow bq$ and
$N \rightarrow b^{-d}N$, we find that the $n$-th GL expansion in $N$ scales as
$b^{n-1-d}$. This means that if we have a local GL theory, all high order
expansions in $N$ are irrelevant in the tree level. However, since the
coefficient $\Gamma_n $ scales as $b^{d-n-1}$ due to the long-ranged
interaction, {\it all} higher order terms gain marginal scaling.
Therefore, there 
are infinitely many marginal operators in this theory, and the Gaussian 
approximation may break down. This is because of the singularities in
$\Gamma_n$ that occurs as $T\rightarrow 0$, which suggests that
 a more careful RG analysis is called for.
  The result of the $\epsilon =4-(d+1)$ expansion in this model
shows that there exists a non-trivial stable fixed point at the AF 
transition\cite{balents98}. In this paper we use the large-$n$ expansion
method  by introducing $n$ fermion species to estimate the 
leading correction to the 
critical exponents and take $n=1$ at the end. 


We come back to Eqs. (\ref{S0}) and 
(\ref{SAF}), and introduce  fictitious $n$ fermion
flavors which couple to the N\'{e}el order parameter symmetrically.
We take the upper momentum cutoff $\Lambda < \Delta_0$ and re-express
the effective action in terms of the fermions coupled to the N\'{e}el
field as follows 
\begin{eqnarray}
S_{\rm eff} &=&\sum_{j=1}^n S_0^j+ 
\int d\omega~ d^dq \big{[} {\bf N}(q)\cdot {\bf N}(-q) ~(
t + q^2 )\big{]} \label{effAct}
 \\ &&~-g~ \sum_j
e^{-i{\bf Q\cdot x}}{\bf N}( x)\cdot
{1\over 2} \psi^{\dag}_{\alpha,j}(x)~\vec{\sigma} ^{\alpha}_{\beta}
\psi^{\beta}_j(x) \Big{\}}~, \nonumber  
\end{eqnarray}
where each fermion species is labeled by the index $j$. We have not included
higher order local expansions in  ${\bf N}$ since more significant non-local
expansions will be perturbatively generated by the long-wavelength fermions.
Here each $S_0^i$ is of the same form as Eq. (\ref{S0}) except that the
momentum and frequency are confined to within a sphere of radius 
$\Lambda \ll \Delta_0$ so that
the fermionic spectrum can be well-approximated by the Dirac spectrum.

\begin{figure}[t]
\epsfxsize=6.5cm \epsfbox{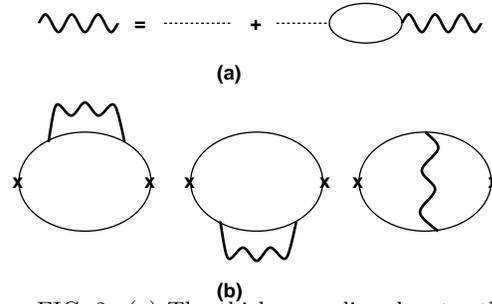}
\caption{(a) The thick wavy line denotes the leading order N\'{e}el propagator 
in the large-$n$ limit, and the thin solid line is the bare fermion 
propagator.  The dashed line is the bare AF exchange coupling
and the bubble is the bare fermionic AF spin susceptibility.
(b) Corrections to the N\'{e}el self-energy of order $O(1/n)$. 
}
\label{bubble}
\end{figure}

Now we take the large-$n$ limit and  estimate the critical
exponents by calculating the N\'{e}el propagator. Note that $n$ is not
the number of order parameter components  but the number of 
fermion species. We find that this is a convenient choice since the
singular GL coefficients in Eq. (\ref{gl}) are ultimately determined by the 
long-wavelength fermions and the coupling $g$ in Eq. (\ref{effAct}).
In this way, we can systematically calculate $g$-independent critical 
exponents as expansions in $1/n$ by the perturbation technique\cite{Ma}.
In Eq. (\ref{gl}), each vertex function $\Gamma_n$ consists of one 
fermionic loop which has a factor of $n$. This means that $1/n$ expansion
is equivalent to loop expansions in the N\'{e}el field---$1/n$ takes the
role of $\hbar$. In practice, it is more convenient to directly work with 
fermions  rather than to integrate them out at the outset to
perform loop expansions in the order parameter field.
Figure \ref{bubble} (a) shows the leading N\'{e}el propagator in the large-$n$
limit, which is $\langle N(q)N(-q)\rangle =[n g^2 |q|/16 +t ]^{-1}$ in
$d=2$, and  $\sim [ |q|^{d-1} +t]^{-1}$ in general dimensions.
Note that the leading non-analyticity is already obtained here. The critical
exponents at the leading order agree with the Gaussian approximation of
Eq. (\ref{gl}). 
In order to estimate the $1/n$ correction to the critical exponents, we
calculate the diagrams shown in Fig. \ref{bubble} (b). 
Here we focus on two spatial dimensions, as it is difficult to carry out 
analytic calculations in general dimensions. 
Setting $t=0$,
we can calculate the correction to $\eta $ by finding the coefficient of
the term $|q| \ln |q|$ in the N\'{e}el self-energy correction.
In $d=2$, since there is fermion self-energy correction $\sim (1/n)~k\ln |k|$,
we may expect a N\'{e}el self-energy
correction of the form $(1/n)~|q| \ln |q|$ on the dimensional ground. 
Explicit calculation shows that
there is no such logarithmic correction from any of the three diagrams in
Fig. \ref{bubble} (b). Therefore, we obtain $\eta = 1 +O(1/n^2)$.
Setting a non-zero $t>0$, we may obtain the correction to $\gamma$
by finding the coefficient of the self-energy correction of the form $t \ln t$,
and we find that $\gamma \approx 1+0.78/n +O(1/n^2)$.

The long-ranged interaction discussed above is dominant only
in the $T\rightarrow 0$ limit, but the finite-temperature crossover 
behavior\cite{millis93,sachdev97} is beyond the scope
of discussion in this report. We merely comment on the classical 
finite-temperature transition for $2<d<3$ where the finite-temperature
N\'{e}el order is assumed to be stabilized by weak interlayer coupling
of cuprates,
assuming that the nodal quasiparticles 
survive at non-zero temperatures. When the temperature is the largest
energy scale of the problem, the non-local interaction of the N\'{e}el
order parameter gives subleading power laws near the transition.
As a result, the leading Gaussian GL expansion is of the form
$\int d^dq ( {\bf q}^2+t) {\bf N}({\bf q})\cdot  {\bf N}({\bf -q})$ 
where $t=(T-T_c)/T_c$. This coincides with the usual classical GL functional,
and therefore we find no distinctive critical behavior in this case.

As pointed out by Hertz\cite{hertz}, AF transitions on nested Fermi surfaces
and superconducting transitions in Fermi liquids are two other examples 
which suffer from the $T\rightarrow 0$ singularities in GL coefficients. 
 In fact, in these two examples, zero-temperature GL expansions 
are more difficult because of $\ln |{\bf q}|$ singularity in the Gaussian 
term. This is related to the fact that the Fermi liquid ground state is always 
unstable in these cases.


We have shown that the existence of a $d$-wave gap in the quasiparticle 
spectrum leads to distinctive critical behavior at zero-temperature
phase transitions, due to the effective long-ranged interaction
of the order parameter. 
We showed that the nodal liquid phase is stable against
an AF instability within a sizable window of AF exchange coupling
and that there exists a non-trivial fixed point at the AF transition.
Assuming that it is possible  to access the quantum critical
regime, critical behavior of thermodynamic quantities 
 would be observed provided that the  distance 
from the quantum critical point $t$ can be controlled. In principle,
$t$ depends on the AF exchange interaction which in turn
is controlled by doping. 
Estimates from two-magnon Raman scattering experiments on cuprates\cite{raman}
show that the effective AF super-exchange coupling $J$ takes a
value of $\sim 125~ \rm meV$
for AF insulators and decreases as doping.
Here we assume that the only effect of doping is to reduce the effective
exchange coupling, and also ignore the possibility of a spin-density-wave
ordering with incommensurate wave-vectors\cite{neutron}.
We expect a scaling form of 
thermodynamic quantities in the quantum critical regime
such as the critical specific heat coefficient, 
$\gamma_V =c_V/T \sim T^{d-1}F(t^{\nu z}/T)$ in terms of a scaling 
function $F(x)$.
From  the scaling relation, we find that $z\nu = z\gamma /(2-\eta )
\approx 1+0.78/n +O(1/n^2)$.
This analysis may provide a check on the notion of
the nodal liquid as the zero-temperature pseudogap phase of the underdoped
cuprates. Also we notice that the $d$-wave like excitation is possibly
ubiquitous even in undoped or underdoped insulating cuprates, as recently 
revealed by the angle-resolved photoemission spectroscopy on an insulating 
parent compound 
$\rm Ca_2CuO_2Cl_2$\cite{Ronning}, and theories on nodal excitations
merit further attention.

 It would  be also interesting to study the effect of nodal quasiparticles 
on the superconductor-insulator
transition in the cuprates. At first glance, however,  the coupling of the 
superconducting order parameter to nodal quasiparticles is suppressed by a 
$d$-wave angular factor of $\cos 2\phi$ near $\phi = \pi (2m+1)/4$ and we 
expect that it is
subleading in the RG sense near the quantum phase transition of the
 $2+1$ dimensional $XY$ type. 

The author would like to thank Alan Dorsey,  Brad Marston, Pierre Ramond, 
Subir Sachdev, and Rob Wickham for helpful discussions, and Matthew Fisher
for pointing out an error in an earlier manuscript.
This work was supported  by the 
National High Magnetic Field Laboratory and by NSF grant DMR 96-28926.

\end{document}